\documentclass[preprint,3p,authoryear,twocolumn]{elsarticle}
\usepackage{graphicx}
\usepackage{amssymb}
\usepackage{lineno}
\begin{document}
\begin{frontmatter}
\title{Two-frequency approach to the theory of atmospheric acoustic-gravity waves}
\author{O.K. Cheremnykh}
\author{E.I. Kryuchkov}
\author{A.K. Fedorenko}
\author{Yu.O. Klymenko$^*$}
\ead{yurkl@ikd.kiev.ua, yurklym@gmail.com}
\cortext[cor1]{Corresponding author.}

\address{Space Research Institute, prosp. Akad. Glushkova 40, build. 4/1, 03187 MSP Kyiv-187, Ukraine}
\begin{abstract}
The propagation of acoustic-gravity waves (AGWs) in the stratified isothermal atmosphere is analyzed using methods of the oscillation theory. It is shown that AGW in the atmosphere can be considered as an oscillatory process occurring at two eigenfrequencies. This consideration makes it possible to explain some of the observed properties of AGWs. The solutions for perturbed hydrodynamic velocity versus time and spectral characteristics are obtained in a real, but not complex, variables.
\end{abstract}
\end{frontmatter}

\section{Introduction}
Since the middle of the last century, the acoustic-gravity waves (AGWs)  in the atmospheres of the Earth and the Sun have been intensively studied both experimentally and theoretically (Hines, 1960; Yeh and Liu, 1974; Nappo, 2002; Sutherland, 2010; Narayanan, 2013). In modern analytical and numerical models of AGWs, including the nonlinear theory, oscillations with acoustic and gravitational frequencies are considered separately (Stenflo et al., 2009; Kaladze et al., 2008; Vadas, 2012). In theoretical models for these waves, it is usually used a representation of perturbed quantities in the form of complex functions. Their substitution into the system of linearized hydrodynamic equations leads to the well-known dispersion equation for AGWs as well as to the ``polarization'' relations between the complex perturbed quantities (Hines, 1960).

In this paper, we show that AGW in the isothermal atmosphere can be considered and studied in the framework of the standard oscillation theory as a system of two-coupled oscillators. In contrast to approaches adopted in AGW linear theory, we will look for wave solutions for physical components of perturbed velocities and displacements in the form of real quantities depending on time, coordinates, spectral characteristics of disturbances and initial conditions. In our opinion, namely real but not complex solutions are more intuitive for comparison of theoretical results with experimental data. As far as we know, such a problem has been not previously solved in the conventional AGW theory.

\section{Basic equations} Dynamics of small disturbances in the stratified isothermal atmosphere is described by well-known linearized hydrodynamic equations (see, for example, Landau, 1969, I):
\begin{equation}\label{1}
\qquad\quad
\begin{array}{lll}
\rho \frac{\partial\vec{V}}{\partial t} =-\nabla p'+\rho'\vec{g},\\
\frac{\partial p'}{\partial t} =-\vec{V}\cdot \nabla p-\gamma p\,\textrm{div}\,\vec{V},\\
\frac{\partial \rho '}{\partial t} =-\vec{V}\cdot \nabla \rho -\rho \,\textrm{div}\,\vec{V}.
\end{array}
\end{equation}
Here, $t$ is the time, $\vec{V}$ is the perturbed hydrodynamic velocity, $p$ is the pressure, $\rho $ is the medium density, and $\gamma $ is the ratio of specific heats. The perturbed quantities of the pressure and the density are indicated by a prime. The equilibrium density $\rho$ entering Eqs.(1) is vertically stratified in a field of gravity, namely
\begin{equation}\label{2}
\qquad\quad\rho \left(z\right)=\rho \left(0\right)\exp \left(-z/H\right).
\end{equation}
Here $H=RT/g$ is the atmospheric scale height, $R$ is the universal gas constant, $T=\textrm{const}$ is the temperature, and $g$ is the acceleration of the gravity.

For further consideration, it is convenient to rewrite Eqs.(1) in terms of the vector of the displacement of medium elementary volume from its equilibrium position:
\begin{equation}\label{3}
\qquad\qquad \vec{\xi }\left(\vec{r},t\right)=\int\limits_{0}^{t}\vec{V}\left(\vec{r},\tau \right) d\tau.
\end{equation}
As a result, we get four equations written in Cartesian coordinates $(x,z)$ with the vertical coordinate $z$:
\begin{equation}\label{4}
\!\!\!\!\!\!\!\!\begin{array}{cc}
\frac{\partial^{2}\xi _{x}}{\partial t^{2}}=-\frac{1}{\rho}\frac{\partial p'}{\partial x},&
\frac{\partial^{2}\xi _{z}}{\partial t^{2}}=-\frac{1}{\rho}\frac{\partial p'}{\partial z}-\frac{\rho'}{\rho}g,\\
\rho'=-\xi_{z}\frac{\partial\rho }{\partial z}-\rho\,\textrm{div}\,\vec{\xi},&
p'=-\xi _{z}\frac{\partial p}{\partial z} -\gamma p \,\textrm{div}\,\vec{\xi }.
\end{array}
\end{equation}
Here $\xi_{x}$ and $\xi_{z}$ are the horizontal and the vertical components of the displacement vector. For obtaining the second pair of Eqs.(4), we integrated over time the equations for the perturbed quantities $p'$and $\rho'$ under assumption that the perturbations are equal to zero at the initial moment of time. This assumption is commonly used to study wave processes in the continuous media (Bateman, 1976; Prist, 1985; Mihalas et al., 2013).

After some algebraic transformations, Eqs.(4) are reduced to system of two second-order linear differential equations written for displacements $\xi_{x} $ and $\xi_{z}$ (Tolstoy, 1963; Cheremnykh et al., 2019):
\begin{equation}\label{5}
\!\!
\begin{array}{c}
\rho\frac{\partial^{2}\xi_{x}}{\partial t^{2} }=-\rho g\frac{\partial\xi_{z}}{\partial x}+\frac{\partial}{\partial x} \left[\rho c_{s}^{2} \left(\frac{\partial\xi_{x}}{\partial x}+\frac{\partial\xi _{z}}{\partial z}\right)\right],\\
\rho\frac{\partial ^{2} \xi _{z} }{\partial t^{2} } =\rho g\frac{\partial \xi _{x} }{\partial x} +\frac{\partial }{\partial z} \left[\rho c_{s}^{2} \left(\frac{\partial \xi _{x} }{\partial x} +\frac{\partial \xi _{z} }{\partial z} \right)\right].
\end{array}
\end{equation}
Here, $c_{s} =\left(\gamma RT\right)^{1/2} $ is the sound velocity. The system of Eqs.(5) describes coupled movements of a medium elementary volume in the vertical and the horizontal directions.

\section{Initial conditions}

To study the displacement vector dependence versus the time with help of Eqs.(5), it is necessary to specify the initial conditions. When obtaining Eqs.(4), it has been taken into account that the perturbed values and the displacement vector have to be zero at the initial moment of the time. Therefore, for displacements $\xi _{x}\left(t,x,z\right)$ and $\xi _{z} \left(t,x,z\right)$ at $t=0$ one can write down conditions
\begin{equation}\label{6}
\qquad\xi _{x} \left(0,x,z\right)=0,\quad \xi _{z} \left(0,x,z\right)=0
\end{equation}
which we will consider as initial ones. The initial conditions for the components of the perturbed velocity $V_{x} =\dot{\xi}_{x} $ and $V_{z}=\dot{\xi }_{z} $ at $t=0$ we represent in form:
\begin{equation}\label{7}
\begin{array}{c}
V_{x}\left(0,x,z\right)=V_{0x} e^{z/2H}\cos\left(\vec{k}\cdot \vec{r}-\phi _{x} \right),\\
V_{z}\left(0,x,z\right)=V_{0z}e^{z/2H}\cos\left(\vec{k}\cdot\vec{r}-\phi _{z}\right)
\end{array}
\end{equation}
following (Landau, 1969, II; Prist, 1982). In Eqs.(7), $\vec{k}\cdot \vec{r}=k_{x} x+k_{z} z$. The exponential factor takes into account vertical density stratification (Hines, 1960), $\phi _{x} $ and $\phi _{z} $ are the oscillation phases for the velocity components. This way we come to the problem of finding solutions to Eqs.(5) supported initial conditions (6) and (7).

\section{Solutions}

Since the components of the displacement vector $\xi _{x} \left(t,x,z\right)$ and $\xi _{z} \left(t,x,z\right)$ equal to zero at initial moment $t=0$ (see Eqs.(6)), it is easily to understand that solutions of Eqs.(5) should be look for in the form of linear combinations
\begin{equation}\label{8}
\!\!\!\!\!\!\!\!\!\!\!\!
\begin{array}{l}
\xi _{x} =e^{z/2H} \sin \omega t\left[a\cos \left(\vec{k}\cdot \vec{r}\right)+b\sin \left(\vec{k}\cdot \vec{r}\right)\right],\\
\xi _{z} =e^{z/2H} \sin \omega t\left[c\cos \left(\vec{k}\cdot \vec{r}\right)+d\sin \left(\vec{k}\cdot \vec{r}\right)\right],
\end{array}
\end{equation}
where $a$, $b$, $c,$ and $d$ are arbitrary constants. Substituting expressions (8) into Eqs.(5) and taking into account the linear independence of trigonometric functions, one can obtain the system of linear equations
\begin{equation}\label{9}
\begin{array}{l}
a\left(\omega^{2}-\omega_{01}^{2}\right)-c\omega _{04}^{2}-d\omega _{03}^{2} =0, \\
b\left(\omega ^{2}-\omega_{01}^{2} \right)+c\omega _{03}^{2}-d\omega_{04}^{2} =0, \\
c\left(\omega ^{2} -\omega _{02}^{2} \right)-a\omega _{04}^{2} +b\omega _{03}^{2} =0,\\
d\left(\omega ^{2} -\omega _{02}^{2} \right)-a\omega _{03}^{2} -b\omega _{04}^{2} =0.
\end{array}
\end{equation}
This system has a nontrivial solution if its discriminant is equaled to zero. It gives dispersion relation
\begin{equation}\label{10}
\qquad\omega _{1,2}^{2} =\frac{1}{2} \left(\omega _{01}^{2} +\omega _{02}^{2} \pm \sqrt{D}\right)
\end{equation}
with
$$D=\left(\omega _{01}^{2} -\omega _{02}^{2} \right)^{2} +4\left(\omega _{03}^{4} +\omega _{04}^{4} \right)$$
where the following notations have been used:
$$
\begin{array}{l}
\omega _{01}^{2} =k_{x}^{2} c_{s}^{2},\qquad \omega_{02}^{2} =\left(k_{z}^{2} +\frac{1}{4H^{2} } \right)c_{s}^{2},\\
\omega _{03}^{2} =\varepsilon \frac{k_{x}}{H}c_{s}^{2}, \quad
\omega _{04}^{2} =k_{x} k_{z} c_{s}^{2},  \quad
\varepsilon =\frac{1}{\gamma } -\frac{1}{2}.
\end{array}
$$
The eigenfrequencies $\omega _{1} $ and $\omega _{2} $ in the theory of acoustic-gravitational waves are usually called ``acoustic'' and ``gravitational'' frequencies, respectively (Hines,1960, Yeh and Liu, 1974).

Substitution of expressions (10) into Eqs.(9) makes its a linearly dependent. Leaving in Eqs.(9) only the last two equations, one can express coefficients $c$ and $d$ through $a$ and $b$ for each of the eigenfrequencies as
\begin{equation}\label{11}
\!\!\!\!\!\!\!\!\! c_{k}=\frac{\left(a_{k}\omega _{04}^{2}-b_{k}\omega _{03}^{2}\right)}{\omega_{k}^{2}-\omega_{02}^{2}},\, d_{k} =\frac{\left(a_{k} \omega _{03}^{2} +b_{k} \omega _{04}^{2} \right)}{\omega _{k}^{2} -\omega _{02}^{2} }, \end{equation}
$k=1,2$.
Then, according to the linear theory of oscillations (Landau and Lifshitz, 1969, I; Arnold, 1989) the general solution of Eqs.(5) has to be written as a sum of two linearly independent solutions (8) with  frequencies (10). Namely,
\begin{equation}\label{12}
\!\!\!\!\!\!\!\!\!\!\!\!\!\!\!\!\!\!\!
\begin{array}{l}
\xi_{x}\!=e^{z/2H}\!\!\sum\limits_{k=1}^{2}\sin\omega _{k}t\!\!\left[a_{k}\!\cos \!\left(\vec{k}\cdot \vec{r}\right)\!\!\!+b_{k} \sin \!\left(\vec{k}\cdot \vec{r}\right)\right]\!,\\
\xi _{z} =e^{z/2H}\sum\limits_{k=1}^{2}\sin \omega _{k} t\left[\frac{a_{k} \omega _{04}^{2} -b_{k} \omega _{03}^{2} }{\omega _{k}^{2} -\omega _{02}^{2} }\!\cos \!\left(\vec{k}\cdot \vec{r}\right)\!\!\!+\right.\\
\left.\qquad\qquad\qquad\qquad\frac{\left(a_{k} \omega _{03}^{2} +b_{k} \omega _{04}^{2} \right)}{\omega _{k}^{2} -\omega _{02}^{2} } \cos \!\left(\vec{k}\cdot \vec{r}\right)\right]\!.
\end{array}
\end{equation}
Here, coefficients $a_{k} $ and $b_{k} $ can be found from initial conditions (7).

Using Eqs.(7), (12) and taking into account that $V_{x} =\dot{\xi }_{x} $ and $V_{z} =\dot{\xi }_{z} $, we obtain  following expressions for perturbed velocity components:
\begin{equation}\label{13}
\!\!\!\!\!\!\!\!\!\!\!\!\!\!\!\!\!
\begin{array}{l}
V_{x}=e^{z/2H}\cos\omega_{1}t\left[\frac{V_{0x}Q_{-} }{2}\cos\!\left(\vec{k}\cdot \vec{r}-\phi_{x} \right)\!+\right.\\
\left.
\frac{V_{0z}\omega _{04}^{2}}{\sqrt{D}} \!\cos\! \left(\vec{k}\cdot \vec{r}-\phi _{z} \right)-
\frac{V_{0z} \omega _{03}^{2}}{\sqrt{D} } \sin \!\left(\vec{k}\cdot \vec{r}-\phi _{z} \right)\right]\!\!+\\
e^{z/2H} \cos \omega _{2}t\left[\frac{V_{0x}Q_{+} }{2} \cos \!\left(\vec{k}\cdot \vec{r}-\phi _{x} \right)-\right.\\
\left.\frac{V_{0z}\omega _{04}^{2} }{\sqrt{D} } \cos \!\left(\vec{k}\cdot \vec{r}-\phi _{z} \right)+\frac{V_{0z}\omega _{03}^{2} }{\sqrt{D} }  \sin\! \left(\vec{k}\cdot \vec{r}-\phi _{z} \right)\right]\!\!,\\
V_{z} =e^{z/2H} \cos \omega _{1} t\left[\frac{V_{0z}Q_{+} }{2} \cos \!\left(\vec{k}\cdot \vec{r}-\phi _{z} \right)\,+\right.\\
\left.\frac{V_{0x}\omega _{04}^{2} }{\sqrt{D} }  \cos\! \left(\vec{k}\cdot \vec{r}-\phi _{x} \right)+\frac{V_{0x}\omega _{03}^{2} }{\sqrt{D} }  \sin \!\left(\vec{k}\cdot \vec{r}-\phi _{x} \right)\right]\!\!+\\
e^{z/2H} \cos \omega _{2} t\left[\frac{V_{0z}Q_{-} }{2} \cos\!\left(\vec{k}\cdot \vec{r}-\phi _{z} \right)-\right.\\
\left.\frac{V_{0x}\omega _{04}^{2} }{\sqrt{D} }  \cos \!\left(\vec{k}\cdot \vec{r}-\phi _{x} \right)-\frac{V_{0x}\omega _{03}^{2} }{\sqrt{D} }  \sin \!\left(\vec{k}\cdot \vec{r}-\phi _{x} \right)\right]\!\!.
\end{array}
\end{equation}
Here
\[\qquad\qquad Q_{\pm}=1\pm\frac{\omega _{02}^{2} -\omega _{01}^{2} }{\sqrt{D} }.\]

Eqs.(13) describe the resulting oscillations of the elementary volume in the isothermal atmosphere under the influence of gravity and pressure gradients. It is seen that this resulting motion is the sum of two harmonic oscillations at eigenfrequencies $\omega _{1} $ and $\omega _{2} $ defined from dispersion relation (10).

\section{Analysis of solutions}

The dimensionless velocity components $\frac{V_{x}}{V_{0x}}$ and $\frac{V_{x}}{V_{0x}}$ versus  dimensionless time $\tau =\omega _{02} t$ are represented in Figs.1-3 for different $k_{x} H$ and $k_{z} H$. It has been assumed that $V_{0x} =V_{0z} =V_{0} $, $\phi _{x} =\phi _{z} =0$, $z=0$, $k_{x} x=\pi /4$, and $\gamma =1.67$. One can see that the oscillations periods and the velocity components amplitudes essentially depend on the values of the quantities $k_{x} H$ and $k_{z} H$.
\begin{figure}[tb]
\includegraphics[width=\columnwidth]{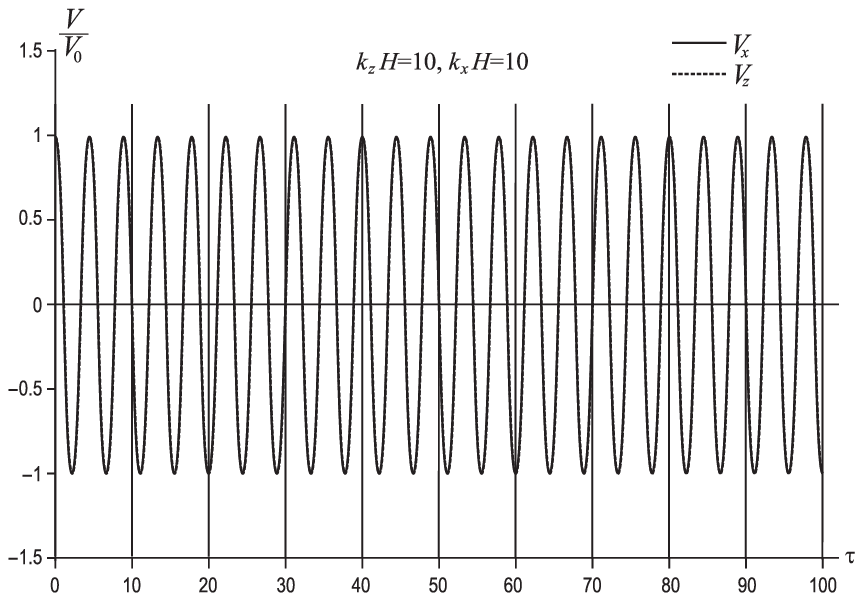}
\caption{Dependencies $\frac{V_{x}}{V_{0}}$ and $\frac{V_{z}}{V_{0}}$ versus $\tau =\omega _{02} t$ derived from Eq.(13). Here, $k_{x}H=k_{z}H=10$.}
\label{fig:1}
\end{figure}
%
\begin{figure}[tb]
\includegraphics[width=\columnwidth]{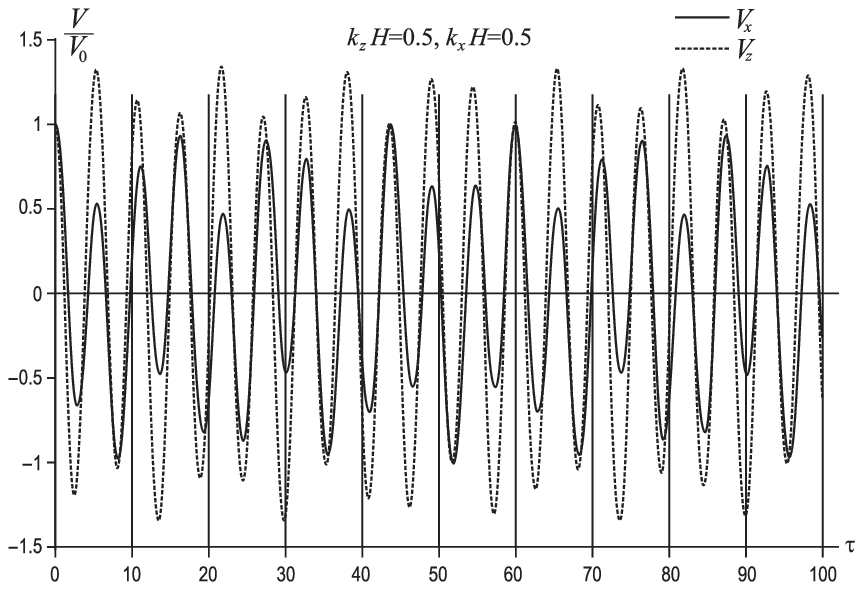}
\caption{Dependencies $\frac{V_{x}}{V_{0}}$ and $\frac{V_{z}}{V_{0}}$ versus $\tau =\omega _{02} t$ for $k_{x}H=k_{z}H=0.5$.}
\label{fig:2}
\end{figure}
%
\begin{figure}[tb]
\includegraphics[width=\columnwidth]{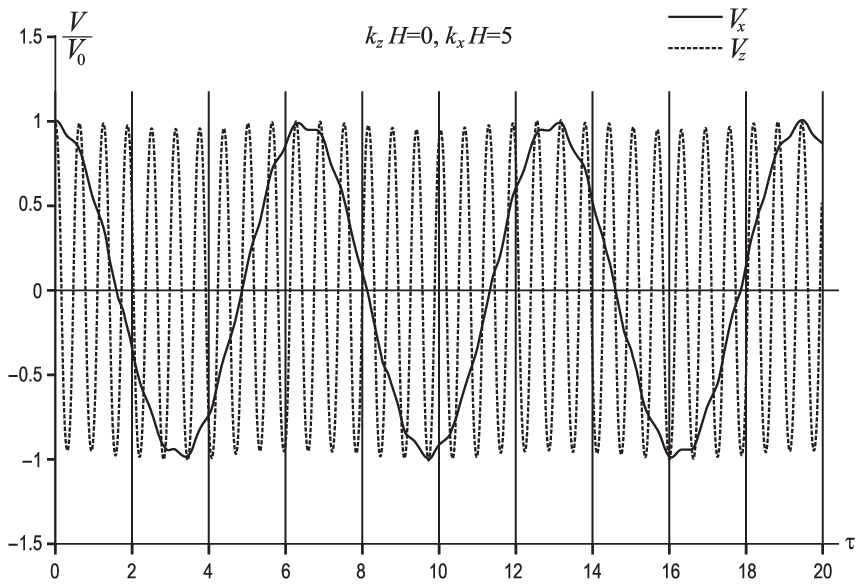}
\caption{Dependencies $\frac{V_{x}}{V_{0}}$ and $\frac{V_{z}}{V_{0}}$ versus $\tau =\omega _{02} t$ for $k_{z} H=0$ and $k_{x} H=5$.}
\label{fig:3}
\end{figure}

It should be noted that in some cases, along with the two-frequency oscillations can be realized the single-frequency oscillation regime. For example, for $k_{z} H\approx k_{x} H>>1$, $V_{0x} =V_{0z} $, and $\phi _{x} =\phi _{z}$ solutions (13) represent high-frequency oscillations
\[\!\!\!\!\!\!\! V_{x} =V_{z} \approx e^{z/2H} V_{0x} \cos \omega _{1} t\cos \left(k_{x} x+k_{z} z-\phi _{x} \right)\]
with frequencies $\omega _{1}^{2} \approx \left(k_{x}^{2} +k_{z}^{2} \right)c_{s}^{2} +\omega _{a}^{2} $ and $\omega _{a}^{2} =c_{s}^{2} /4H^{2} $. This result confirms the behavior of the perturbed velocity components shown in Fig. 1.

The transition to one-frequency oscillations can also be seen in the example of evanescent modes considered in the work (Cheremnykh et al., 2019). At the point $k_{z} =0$, $k_{x} ={1 \mathord{\left/{\vphantom{1 2H}}\right.\kern-\nulldelimiterspace} 2H} $ these modes are realized with frequencies $\omega _{1} =\sqrt{g/2H} $ and $\omega _{2} =\sqrt{g\left(\gamma -1\right)/2H} $. In this case, assuming that $V_{0x} =V_{0z} =V_{0} $ and $\phi _{x} =0$, $\phi _{z} =\pi /2$, we obtain from relations (13)
\[\qquad V_{x} =e^{z/2H} V_{0} \cos \omega _{1} t\sin \left(x/2H\right),\]
\[\qquad V_{z} =e^{z/2H} V_{0} \cos \omega _{1} t\cos \left(x/2H\right).\]
When $\phi _{x} =0$ and $\phi _{z} =-\pi /2$ Eqs.(13) transforms to relations
\[\qquad V_{x} =-e^{z/2H} V_{0} \cos \omega _{2} t\sin \left(x/2H\right),\]
\[\qquad V_{z} =e^{z/2H} V_{0} \cos \omega _{2} t\cos \left(x/2H\right).\]
It is easy to see that $\textrm{div}\,\vec{V}=\frac{\partial V_{x} }{\partial x} +\frac{\partial V_{z} }{\partial z} =0$ in the first case, and $\textrm{div}\,\vec{V}=\frac{V_{z} }{H} $ in the second one, which exactly corresponds to the approximations made in the work (Cheremnykh et al., 2019).

For disturbances of velocity components (13) we obtain the following solutions occurring at $V_{0x} =V_{0z} =V_{0} $ and $\phi _{x} =\phi _{z} =0$:
\begin{equation}\label{14}
\qquad V_{x,z} =V_{x,z} \left(t\right)\cos \left(\omega _{+} t+\phi _{x,z} \left(t\right)\right) \end{equation}
where
\[\!\!\!\!\!\!\!\! V_{x} \left(t\right)=V_{z} \left(t\right)=\left[\frac{V_{0}^{2} }{2} \left(1+\cos \left(x/H\right)\cos 2\omega _{-} t\right)\right]^{1/2} ,\]
\[\qquad \tan \phi _{x,z} =\mp \tan \left(x/2H\right)\tan \omega _{-} t,\]
\[\qquad \omega _{\pm } =\frac{\omega _{0} }{2} \left[\left(1+2\varepsilon \right)^{1/2} \pm \left(1-2\varepsilon \right)^{1/2} \right].\]
From solution (14) it follows that the behavior of the velocity components essentially depends on the point at which the oscillations are considered. For example, at $x=0$ we obtain the oscillations in the form of beats
\[\qquad\quad   V_{x,z} =V_{0} \left(t\right)\cos \left(\omega _{+} t\right)\cos \left(\omega _{-} t\right).\]
At points $x=\pi H/2$, Eq.(14) describes harmonic oscillations
$$
\begin{array}{l}
V_{x} =\frac{V_{0} }{\sqrt{2} } \cos \left[\omega _{0} \left(1-\varepsilon \right)t\right],\\
V_{z} =\frac{V_{0} }{\sqrt{2} } \cos \left[\omega _{0} \left(1+\varepsilon \right)t\right]
\end{array}
$$
with different but very close frequencies.

\section{Comparison with experimental data}

Analysis of satellite measurements of AGWs indicates some inconsistencies with well-known theory results (Hines, 1960). Generally, the observed wave trains are morphologically similar to beats (oscillation with close frequencies). They have typical scale\textbf{ }$k_{x} H\approx 0.5$ and propagate almost horizontally with $k_{z} H\to 0$ (Fedorenko et al., 2015). This may indicate in favor of the two-frequency oscillation regime. Moreover, for AGWs with $k_{z} \to 0$ there are almost in-phase oscillations of velocity components $V_{x} $ and $V_{z} $ (Fedorenko, 2013), although there must be phase shift near $\pi /2$ (Hines, 1960). In AGWs over the polar caps of the Earth, the oscillations of the velocity vertical component occur in the phase with the perturbed density (Fedorenko, 2010, 2013), which is also inconsistent with theory (Hines, 1960). These differences can be explained in the framework of the proposed two-frequency approach.

\section{Main results}

For the first time, atmospheric AGWs are analyzed based on two-frequency approach using the methods of classical oscillation theory. It is shown that the propagation of AGWs in the isothermal atmosphere can be considered as the oscillatory process occurring simultaneously at two eigenfrequencies. In some particular cases, the two-frequency oscillatory regime transforms into the usual single-frequency one. It is obtained the real (not complex) solutions describing the change in perturbed velocity as the function of time, spectral characteristics, and initial conditions. The results obtained allow us to explain some features of satellite observations that cannot be interpreted in the framework of the classical AGW theory.

The work was supported of the Targeted Comprehensive Program of the NAS of Ukraine on Space Research (2018-2022).

\end{document}